\documentclass[]{lni}
\begin{document}
\title{Modellieren mit \textsc{Heraklit}}
\subtitle{Prinzipien und Fallstudie} 
\author[Peter Fettke \and Wolfgang Reisig]
{Peter Fettke\footnote{Saarland University, Saarbr\"ucken, Germany, und German Research Center for Artificial Intelligence (DFKI), Saarbr\"ucken, Germany, \email{peter.fettke@dfki.de}} \and
Wolfgang Reisig\footnote{Humboldt-Universität zu Berlin, Berlin, Germany, \email{reisig@informatik.hu-berlin.de}}}
\startpage{1} 
\editor{M. Riebisch, M. Tropmann-Frick} 
\booktitle{Modellierung 2022} 
\yearofpublication{2022}
\maketitle
\begin{abstract}
\textsc{Heraklit} ist ein laufendes Forschungsprogramm und Entwicklungsprojekt mit dem Ziel der Schaffung einer Infrastruktur zur Modellierung großer, rechnerintegrierter Systeme. Wir diskutieren die zentralen Anforderungen an solche Modelle (Hierarchien, Nutzersicht, Überführung informaler in formale Ideen, schematische Modelle, gleichrangiger Umgang mit digitalisierten und personengebundenen Prozessen) und erläutern, wie \textsc{Heraklit} diese Anforderungen unterstützt. Eine Fallstudie zeigt die Nutzbarkeit von \textsc{Heraklit} in der Praxis.
\end{abstract}


\section{Einleitung}

Informatik hat zwei Gesichter: einerseits die technischen Grundlagen mit den formalen Konzepten zu ihrer Nutzung, andererseits die Anwendung. \textit{Dijkstra} hat vielfach vorgeschlagen, die formale und informelle Sicht zu trennen und zwischen ihnen eine Mauer (engl. „firewall“) zu errichten \cite{dijkstra:firewallcacm}. Begründung: Das  „correctness problem“ des Informatikers verlangt ganz andere Herangehensweisen als das „pleasantness problem“ des Anwenders. Modellierung findet in diesem Bild auf der einen oder anderen Seite dieser Mauer statt.

Demgegenüber schlagen wir hier vor, Modellierung als eine Tätigkeit zu verstehen, die einen möglichst bruchlosen Übergang zwischen formal formulierten und lebensweltlichen  Sachverhalten erlaubt. Modelle verknüpfen Anwendung und Technik und beruhen auf denselben Grundlagen.

In diesem Beitrag beschreiben und motivieren wir zunächst im zweiten Abschnitt Anforderungen, die heutzutage an die Modellierung rechnerintegrierter Systeme gestellt werden. Kapitel 3 stellt dann die Konzepte vor, mit denen \textsc{Heraklit} \cite{fettke2021handbook, fettke2021modelling} diese Anforderungen bewältigt. Eine umfangreiche Fallstudie zeigt im vierten Kapitel, dass die Erfüllung aller Anforderungen in einem integrierten Framework tatsächlich gelingt. Schließlich wird im fünften Kapitel diskutiert, in welchem Umfang andere Frameworks die Anforderungen aus Kapitel 2 entsprechen.

\section{Anforderungen an die Modellierung rechnerintegrierter Systeme}

Der Herstellung eines komplexen rechnerintegrierten Systems ist immer ein Planungsprozess vorgeschaltet, in dem die Architektur, Funktionsweise, beabsichtigten Effekte etc. des Produktes formuliert werden. Zentrales Werkzeug und Hilfsmittel dafür sind Modelle. Auch ein gegebenes komplexes System kann man besser verstehen und analysieren, indem man es modelliert. Ein Modell betont einige Aspekte des Systems, oft mit graphischen, standardisierten Darstellungen. 
Ein gutes Modell ist nicht nur anschaulicher und besser verständlich als eine umgangssprachliche Darstellung; es lässt sich auch vielfältig  nutzen:  man kann Funktionalität und Performanz-Engpässe erkennen, Kosten abschätzen,  Korrektheit  gegenüber  einer  Spezifikation  nachweisen,  Parameter optimieren und vieles andere mehr.

Moderne und zukünftige digitale Systeme, cyber-physikalische Systeme, digitale informationsbasierte Infrastrukturen, das \textit{Internet of people, things and services} etc., werden zunehmend komplexer und wachsen immer mehr zusammen. Folglich sind bei der Gestaltung rechnerintegrierter Systeme vielfältige Herausforderungen zu überwinden. Nötig sind dafür Frameworks, die insbesondere für große Systeme wirklich Nutzen bringen. Dafür ist eine Reihe von Aspekten besonders wichtig:


\textit{Komposition und hierarchische Verfeinerung lokaler Komponenten:} Was bedeutet es konzeptuell, dass rechnerintegrierte Systeme „groß“ sind? Es bedeutet zunächst einmal, dass einige Konzepte, die für kleine Systeme durchaus verwendet werden können, nun nicht mehr praktikabel sind. Am augenfälligsten gilt das für globale Zustände und globale Schritte. Und es  bedeutet, dass ein großes System im Allgemeinen nicht amorph, gleichförmig gestaltet ist, sondern in irgendeiner Form komponiert ist aus kleineren, mehr oder weniger selbständig agierenden Systemen. Komposition und Verfeinerung sind also fundamentale Konzepte zur Modellierung großer Systeme.

\textit{Modelle aus Sicht der Anwender:} Es gibt mehrere Weisen, Informatik im Allgemeinen und Modellierung in der Informatik im Besonderen zu fassen. Häufig wird der Zugang  über  die  theoretische  Informatik  gewählt,  mit  der Betrachtung von Zeichenketten und ihrer Transformation. Beliebt ist auch  der Zugang von der technischen Seite mit digitalen Schaltungen und ihrer Abstraktion, hin zu Software, Datenbanken, etc. Zum Verständnis großer Systeme besonders  nützlich  erscheint aber  der  Zugang  aus  der  Sicht  der  Anwender, Nutzer, Betreiber und der Zwecke, für die große Systeme betrieben werden.

\textit{Systematische Überführung informeller, intuitiver Ideen zu einem gegebenen oder intendierten System in ein formales Modell:} Dafür bietet die Informatik strukturierte Konzepte. Allerdings münden sie zumeist in Programme, nicht in formale Modelle.  

\textit{Parametrisierte, schematische Modellierung:} Oft soll nicht nur ein einziges System modelliert werden, sondern eine Menge gleichartig strukturierter und verhaltensähnlicher Modelle.

\textit{Integrierter, einheitlicher, gleichrangiger Umgang mit digitalisierten und lebensweltlichen Prozessen:} Ein Beispiel ist in einem Betrieb die automatisierte digitale Personalverwaltung, zusammen mit der Belehrung eines Arbeitnehmers  über  gefährliche  Güter  und  dessen  Bestätigung  durch seine Unterschrift. Neben symbolischen, implementierten Konzepten muss ein Modell also auch pragmatische, nicht zur Implementierung vorgesehene Schritte darstellen können. 


\section{Von \textsc{Heraklit} verwendete Konzepte}
Informatik wird üblicherweise von der Rechentechnik her oder vom Symbolischen her aufgefasst, also aus der Sicht der technischen Informatik, oder der klassischen theoretischen Informatik. Demgegenüber betrachtet \textsc{Heraklit} die Informatik aus einer dritten Sicht, der Sicht der Anwender. Für \textsc{Heraklit} zentral ist die holistische Auffassung von Systemen und ihren Modellen: Es reicht nicht, einzelne Anforderungen an das Systemverhalten, Verfahrensanleitungen für Anwender, Datensammlungen, Algorithmen, Softwarekomponenten, Hardwarekonzepte etc. isoliert zu betrachten, zu verstehen und zu verwenden. Vielmehr muss ihr Zusammenspiel verstanden werden. Dafür braucht man ein Modell, das alle diese Aspekte integriert. \textsc{Heraklit} ist auf diese Anforderungen an Modelle zugeschnitten.

\subsection{Konzeptionelle Sicht}
\textsc{Heraklit} unterstützt die Modellierung von Systemen, die aus Teilsystemen zusammengesetzt sind, und die Hierarchien bilden, in denen konkrete Produkte, Maschinen, Menschen, Dokumente, Datenspeicher etc. kombiniert, separiert, transformiert, berechnet, bearbeitet, transportiert, erzeugt, vernichtet werden. \textsc{Heraklit}-Modelle können  abstrakt,  detailliert,  schematisch  sein;  Daten, Gegenstände, Schritte zur Beschreibung dynamischen Verhaltens sind kleinteilig, aber auch umfassend formulierbar. Konkrete und abstrakte Datenstrukturen, sowie die Beschreibung von Verhalten werden integriert und aufeinander bezogen modelliert.  \textsc{Heraklit} bietet  dafür  aufeinander  abgestimmte  Ausdrucksmittel, die den Modellierer bei der Formulierung seiner Ideen unterstützen. Der Modellierer wird nicht eingeschränkt;  bei  der  Wahl  des  Abstraktionsgrades  und  der Art der Komposition von Modulen hat er inhaltlich alle Freiheiten; \textsc{Heraklit} schlägt aber konzeptionell vereinheitlichende Darstellungen vor. \textsc{Heraklit} integriert neue Konzepte mit bewährten, tiefliegend motivierten Konzepten der Modellierung von Software, Geschäftsprozessen und anderen Systemen. Zudem bietet \textsc{Heraklit} ein Konzept zur Abstraktion konkreter Daten, und damit die symbolische Modellierung großer Klassen  von  Systemen,  die  ähnlich  aufgebaut sind und sich ähnlich verhalten. 
 
Mit seinen wenigen, aber ausdrucksstarken Konzepten sind \textsc{Heraklit}-Modelle:
\begin{itemize}
\item für den Entwickler besser beherrschbar,
\item für den Nutzer leicht verständlich,
\item weniger fehleranfällig und leichter verifizierbar,
\item einfacher änderbar, sowie
\item schneller herstellbar und kostengünstiger als andere Methoden, insbesondere auch für wirklich große Systeme.
\end{itemize}

\subsection{Technische Sicht}
Technisch und formal neu ist die Integration von Architektur, statischen und dynamischen Konzepten. Dafür kombiniert \textsc{Heraklit} bewährte mathematisch  basierte  und  intuitiv  leicht verständliche Konzepte wie Prädikatenlogik, abstrakte Datentypen und Petrinetze, die auch bisher schon zur Spezifikation von Systemen verwendet werden; \textsc{Heraklit} kombiniert sie neu und ergänzt sie um den Kompositionskalkül. Das \textsc{Heraklit}-Framework

\begin{itemize}
\item unterstützt  die hierarchische  Partitionierung  eines  großen  Systems in Module;
\item kann technisch einfach, aber inhaltlich flexibel und ausdrucksstark Module zu großen Systemen komponieren;
\item beschreibt diskrete Schritte in großen Systemen lokal konzentriert in einzelnen Modulen in frei gewähltem Detaillierungsgrad;
\item stellt alle Arten von Modulen integriert mit den gleichen Konzepten dar, egal ob sie zur Implementierung vorgesehen sind oder nicht;
\item repräsentiert Daten auf frei gewählter Abstraktionsstufe;
\item berücksichtig Datenabhängigkeiten im Kontrollfluss;
\item kann von konkreten Daten abstrahieren, um verhaltensgleiche Instanziierungen in einem Schema zu fassen;
\item kann Modelle generieren, die skalierbar, systematisch, änderbar und erweiterbar sind;
\item unterstützt den Nachweis, dass ein Modell gewünschte Eigenschaften hat.
\end{itemize}

Mit den beschriebenen technischen Konzepten können Aspekte diskreter Systeme modelliert werden, wie es keine andere Modellierungstechnik ermöglicht: 

\textit{Module unterschiedlicher Hierarchiestufen können komponiert werden.} Beispielsweise kann in einem \textsc{Heraklit}-Modell ein Modul eine Systemkomponente auf feinster, operationeller Stufe modellieren; zugleich werden seine benachbarten Module nur mit ihrer Schnittstelle, ihrem Namen und ggf. Teilen ihrer inneren Struktur repräsentiert.

\textit{Lebensweltliche und abstrakte Objekte sind gleichrangige Elemente formaler Argumentation.} \textsc{Heraklit} folgt der Prädikatenlogik, und fasst lebensweltliche und formale Objekte gleichrangig als Interpretationen von Termen einer Signatur (eines Alphabetes mit getypten Symbolen) auf. 

\textit{Verhaltensmodelle können parametrisiert werden.} Jede Interpretation der symbolischen Beschriftungen eines schematischen \textsc{Heraklit}-Moduls generiert ein eigenes Systemmodell. Beispiel: Ein schematisches Modul beschreibt die Prinzipien der Geschäftsprozesse aller Filialen eines Bekleidungsunternehmens; jede Interpretation beschreibt eine konkrete Filiale. 

\textit{Der Umfang von Modellen wächst linear in der Größe des modellierten Systems.} Das gilt insbesondere auch bei der Modellierung dynamischen Verhaltens. Erreicht wird dies mit dem konsequenten Verzicht auf globale Konstruktionen (beispielsweise globale Zustände) bei der Bildung einzelner Abläufe, mit einem lokal beschränkten Kompositionsoperator, etc.

\textit{Bewährte Verifikationstechniken werden übernommen.} Petrinetze bieten eine Vielzahl effizienter rechnergestützter Verifikationstechniken. Die wichtigsten, insbesondere Platz- und Transitions-Invarianten von Petrinetzen, funktionieren auch auf symbolischer {(Signa\-tur-)} Ebene und – entsprechend ausdrucksstärker – für die einzelnen Interpretationen einer Signatur. Eine Verifikation auf der Basis des \textit{model checking} ist für einzelne Interpretationen einer Signatur möglich; dafür gibt es effiziente Verfahren. Ein großes, offenes Feld ist die kompositionale Verifikation: Eigenschaften eines komponierten Systems werden aus Eigenschaften der komponierten Module abgeleitet.

\textit{Ein einzelner Ablauf kann verteilt sein.} In einem Ablauf eines großen Systems treten Ereignisse oft unabhängig voneinander ein. Beispiel: In einer Verkaufsfiliale interagieren die Mitarbeiter unabhängig voneinander mit einzelnen Kunden; sie werden nur gelegentlich synchronisiert beim Zugang zu knappen Ressourcen, beispielsweise beim Bezahlen an der Registrierkasse. Viele Frameworks verstehen Unabhängigkeit als „beliebige Reihenfolge“; das führt zu exponentiell vielen vermeintlich verschiedenen „sequentiellen“ Abläufen. \textsc{Heraklit} modelliert Unabhängigkeit von Ereignissen in Abläufen explizit. Das führt zu einem klaren Begriff von Nichtdeterminismus und zu einem Theorem, nachdem die Abläufe eines komponierten Systems aus den Abläufen der Komponenten ableitbar sind.

\textit{\textsc{Heraklit} ist intuitiv einfach.} \textsc{Heraklit} verwendet intuitiv einfache Basis-Konzepte: Prädikatenlogik ist vielen Anwendern aus anderen Kontexten ohnehin geläufig. Auch Petrinetze sind weit verbreitet. Der Kompositionskalkül ist an kleinen Beispielen unmittelbar einsichtig. Die graphischen Ausdrucksmittel von Petrinetzen und des \textit{composition calculus} ergänzen sich harmonisch.

\subsection{Die drei Dimensionen von \textsc{Heraklit}} 
\textsc{Heraklit} unterscheidet drei Dimensionen, die drei unterschiedliche, aber integrierte Aspekte der Modellierung kennzeichnen: 


\textit{Architektur:} Ein \textsc{Heraklit}-Modell besteht aus \textit{Modulen}. Jedes Modul hat ein beliebig gestaltetes Inneres; insbesondere ist die Abstraktionsebene der Darstellung frei wählbar. Die Schnittstelle eines Moduls enthält beliebige gelabelte Elemente. Komposition von Modulen ist technisch einfach und immer assoziativ. Formale Grundlage dieses Architekturprinzips ist der Kompositionskalkül aus \cite{reisig2019associative}.

\textit{Statische Aspekte:} Für den Umgang mit Daten und Datenstrukturen greift \textsc{Heraklit} auf abstrakte Datentypen und algebraische Spezifikationen zurück, wie sie sich in der Informatik von Anfang an bewährt haben und in Spezifikationssprachen wie \textit{VDM} \cite{jones1990VDM} und \textit{Z} \cite{bowen2001Z} seit langem verwendet werden \cite{sanella20212algebraic}. Symbolische Darstellungen (Terme einer Signatur) werden wie in der Prädikatenlogik verwendet. Zu jedem Modul gehört eine Signatur, also eine Menge von Symbolen für Mengen, Konstanten und Funktionen, aus denen zusammen mit Variablen dann Terme gebildet werden. Eine Signatur, und damit ihre Terme, können in ganz unterschiedlichen Lebenswelten instanziiert werden. 

\textit{Dynamische Aspekte:} Ein Modul zur Beschreibung von Verhalten enthält in seinem Inneren ein Petrinetz \cite{petri1977non_sequential}. Dabei ist jeder Platz des Petrinetzes ein prädikatenlogisches Prädikat, jeder Pfeil ist mit einem Term der Signatur des Moduls beschriftet, und jede Transition mit einer Bedingung. Für eine gegebene Instanziierung der Signatur kann die Menge der Objekte, auf die das Prädikat zutrifft, durch den Eintritt eines Ereignisses wachsen oder schrumpfen. Im Einzelnen wird das durch Terme an den Pfeilen des Petrinetzes beschrieben. Damit kann Verhalten abstrakt auf schematischer Ebene, aber auch konkret für eine „gemeinte“ Instanziierung dargestellt werden. 

\section{Fallstudie}
Gegeben sei ein Restaurant-Betrieb mit mehreren Filialen. Alle Filialen sind nach demselben Schema aufgebaut und verhalten sich nach demselben Muster; sie unterscheiden sich aber in einigen Einzelheiten. Modelliert wird das Schema aller Filialen, ein Beispiel einer Filiale, und ein konkreter Ablauf in dieser Filiale.

\subsection{Die  Architektur der Filialen}
Grundlegendes Konzept der Architektur der Filialen sind Module, wie sie in Abschnitt 3.2 generell für \textsc{Heraklit}-Modelle skizziert wurden. Im Inneren der hier verwendeten Module liegen Petrinetze in verschiedenen Ausprägungen. Generell hat ein Modul $M$ zwei Schnittstellen, die linke Schnittstelle $^\ast M$ und die rechte Schnittstelle $M ^\ast$. Eine Schnittstelle kann Petrinetz-Plätze und -Transitionen enthalten. Graphisch wird ein Modul rechteckig dargestellt, mit den Elementen der linken Schnittstelle auf dem linken oder oberen Rand des Rechtecks und den Elementen der rechten Schnittstelle auf dem rechten oder unteren Rand.

\begin{figure}[!tb]
\centering
\begin{minipage}{.5\textwidth}
\centering
\includegraphics[scale=.20]{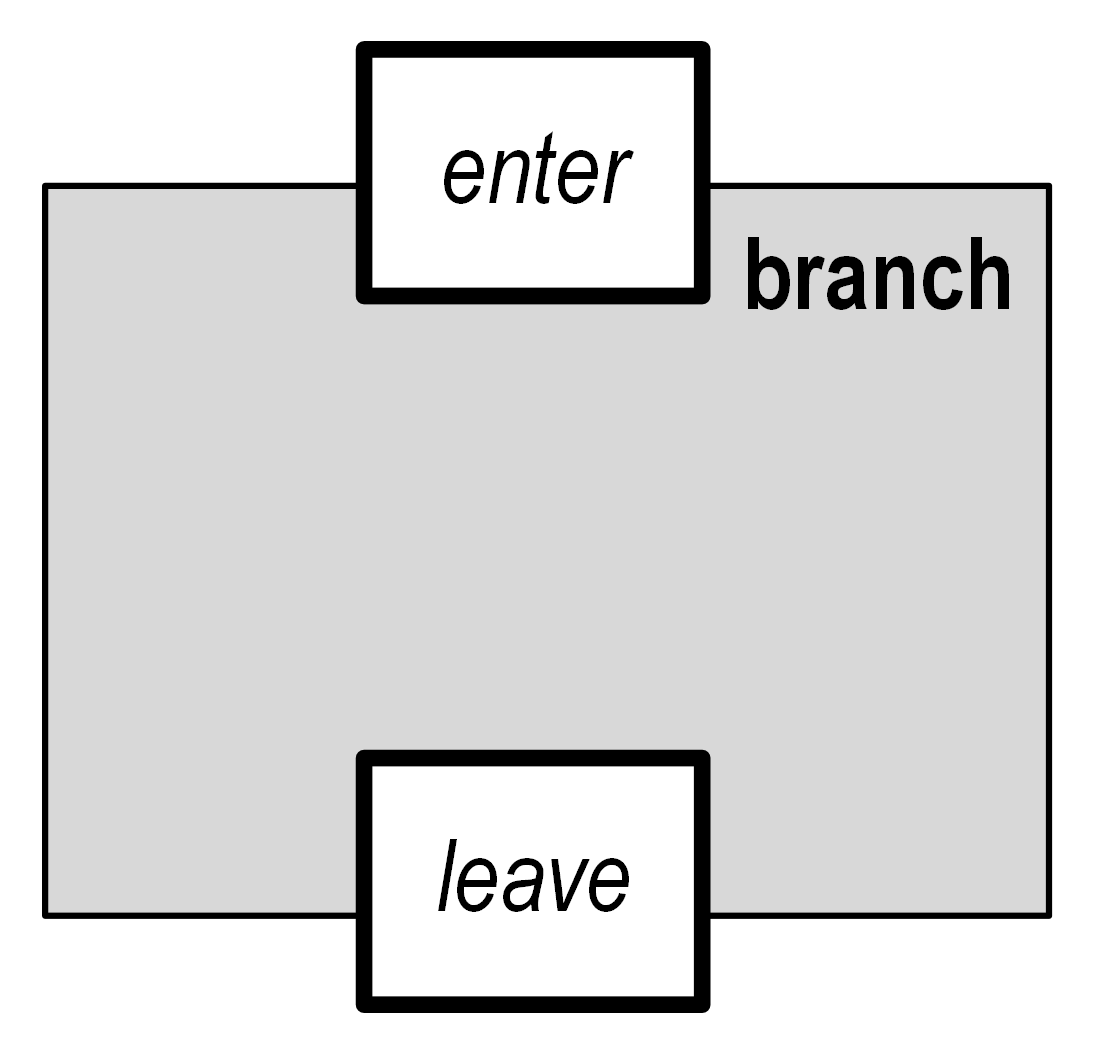}
\caption{abstrakte Sicht auf eine Filiale}
\label{fig:abstrakte_Sicht}
\end{minipage}%
\begin{minipage}{.5\textwidth}
\centering
\includegraphics[scale=.20]{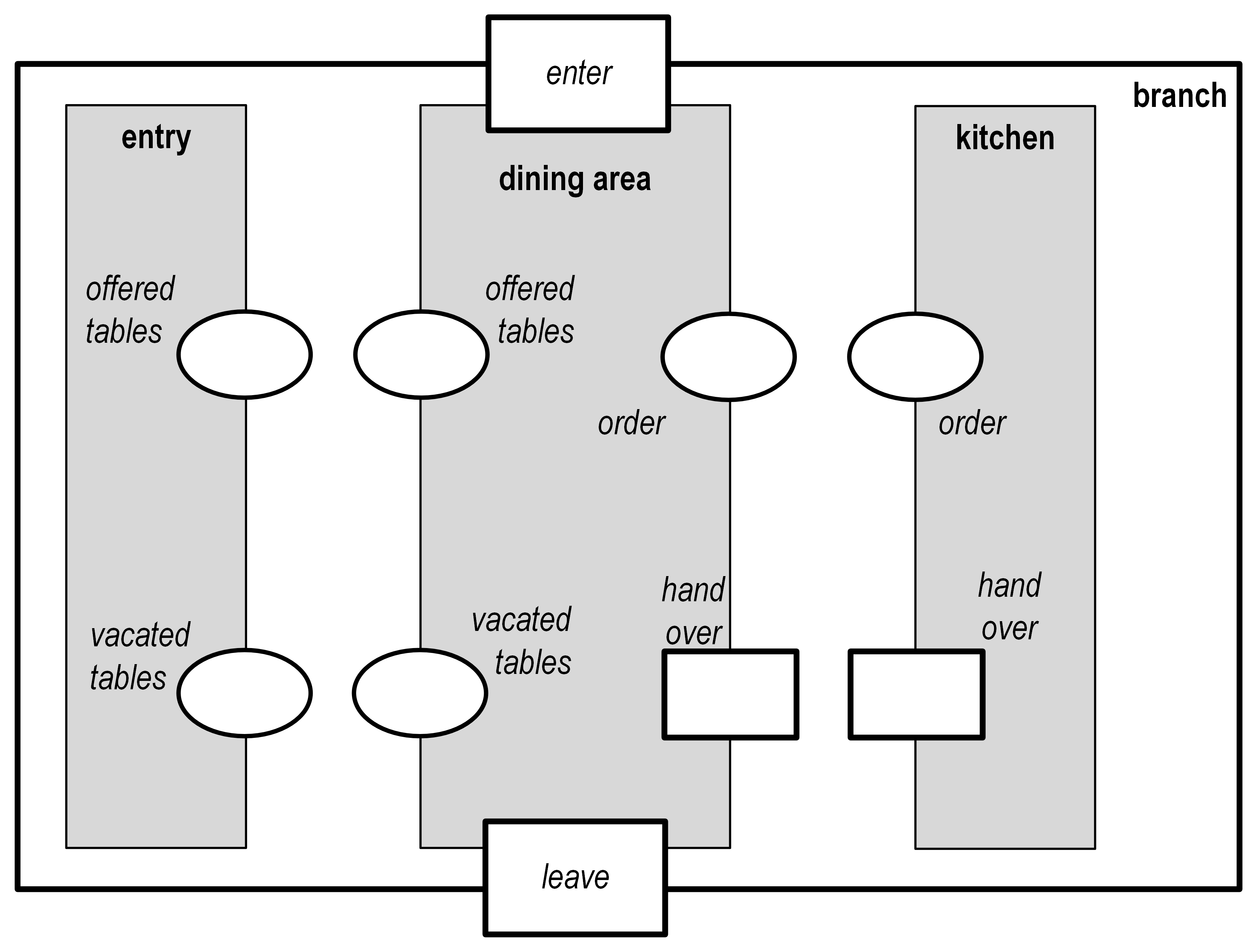}
\caption{eine Filiale, bestehend aus drei Modulen}
\label{fig:drei_Module}
\end{minipage}
\end{figure}

Abb.~\ref{fig:abstrakte_Sicht} zeigt das Schema für die Filialen in größtmöglicher Abstraktion als ein \textsc{Heraklit}-Modul, \textit{branch}. Modelliert werden zwei Aktivitäten der Restaurant-Gäste: das Betreten und das Verlassen einer Filiale; technisch als Petrinetz-Transitionen in $^\ast \textit{branch}$ und $\textit{branch} ^\ast$ mit den Beschriftungen \textit{enter} bzw. \textit{leave}.

Abb.~\ref{fig:drei_Module} zeigt, dass jede Filiale drei Komponenten hat: den Eingang, den Gastraum und die Küche. Diese Komponenten werden wiederum als Module modelliert. Die linke Schnittstelle des Eingangs ist leer, die rechte Schnittstelle enthält zwei Petrinetz-Plätze. Die Beschriftungen dieser beiden Plätze stimmen mit den Beschriftungen der Plätze der linken Schnittstelle des Gastraums überein; gleich beschriftete Plätze werden später beim Komponieren der beiden Module miteinander verschmolzen. Die beiden Transitionen der Schnittstellen des \textit{branch}-Moduls liegen auch in den Schnittstellen des Gastraums. Die rechte Schnittstelle des Gastraums enthält einen Platz und eine Transition; entsprechend beschriftete Elemente enthält die linke Schnittstelle des Küchen-Moduls.

Die Beschriftungen der Schnittstellen-Elemente weisen auf das Verhalten und die Funktionalität der Module hin: am Eingang wird jedem Gast ein freier Tisch zugewiesen. Im Gastraum kann der Gast seine Bestellung aufgeben, später das bestellte Gericht entgegennehmen und an seinem Tisch verspeisen.

\subsection{Statische Komponenten der Filialen}

In jeder Filiale spielen vier Mengen eine zentrale Rolle: die Tische, die Gäste, die Speisekarte mit ihren einzelnen Einträgen und die zubereiteten Gerichte. In jeder Filiale kann jede dieser Mengen jeweils aus anderen Elementen bestehen. Beispielsweise arbeitet eine kleine Filiale mit weniger Tischen und einer anderen Auswahl auf der Speisekarte als eine große Filiale.  

\begin{figure}[!tb]
\centering
\begin{minipage}{.5\textwidth}
\centering
\includegraphics[scale=.25]{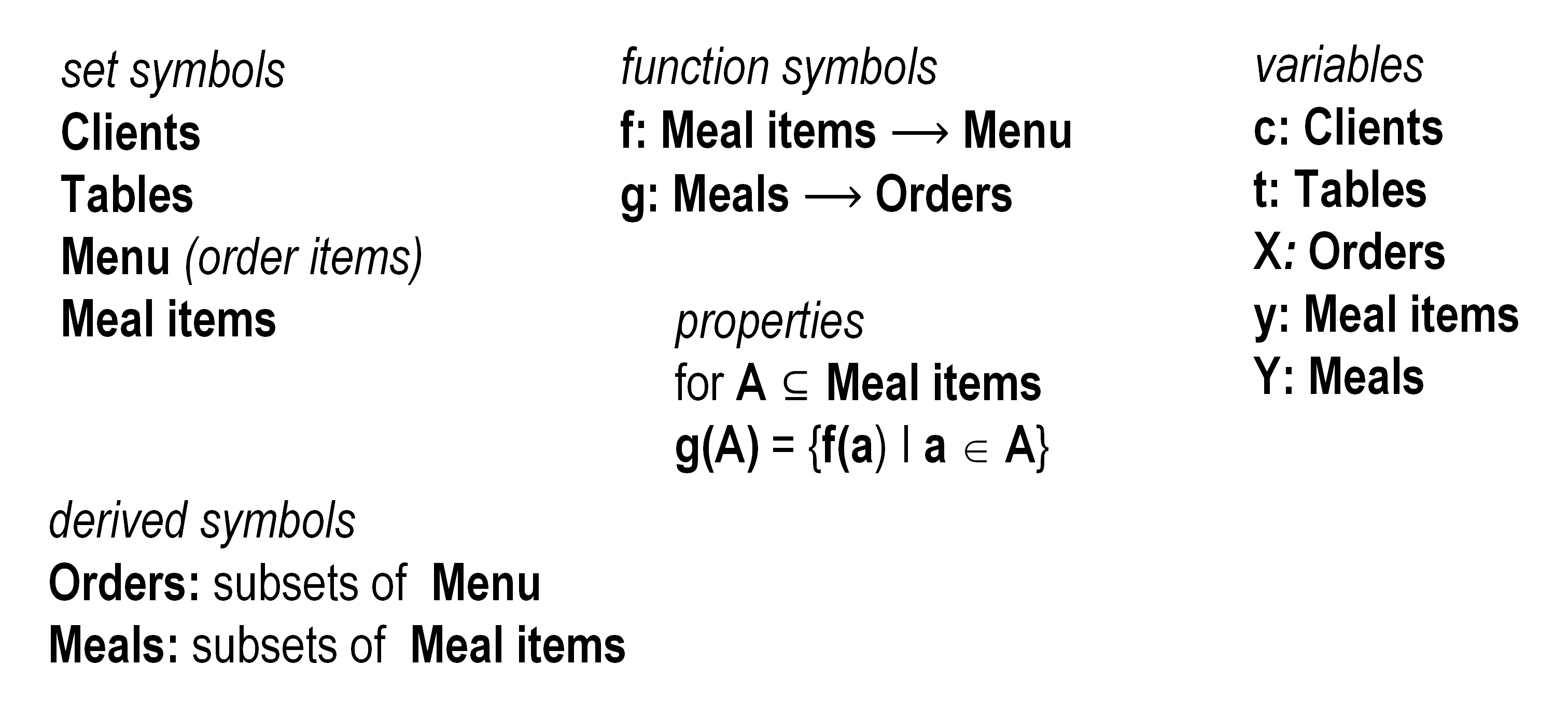}
\caption{Signatur $\Sigma_0$}
\label{fig:Signatur}
\end{minipage}%
\begin{minipage}{.5\textwidth}
\centering
\includegraphics[scale=.25]{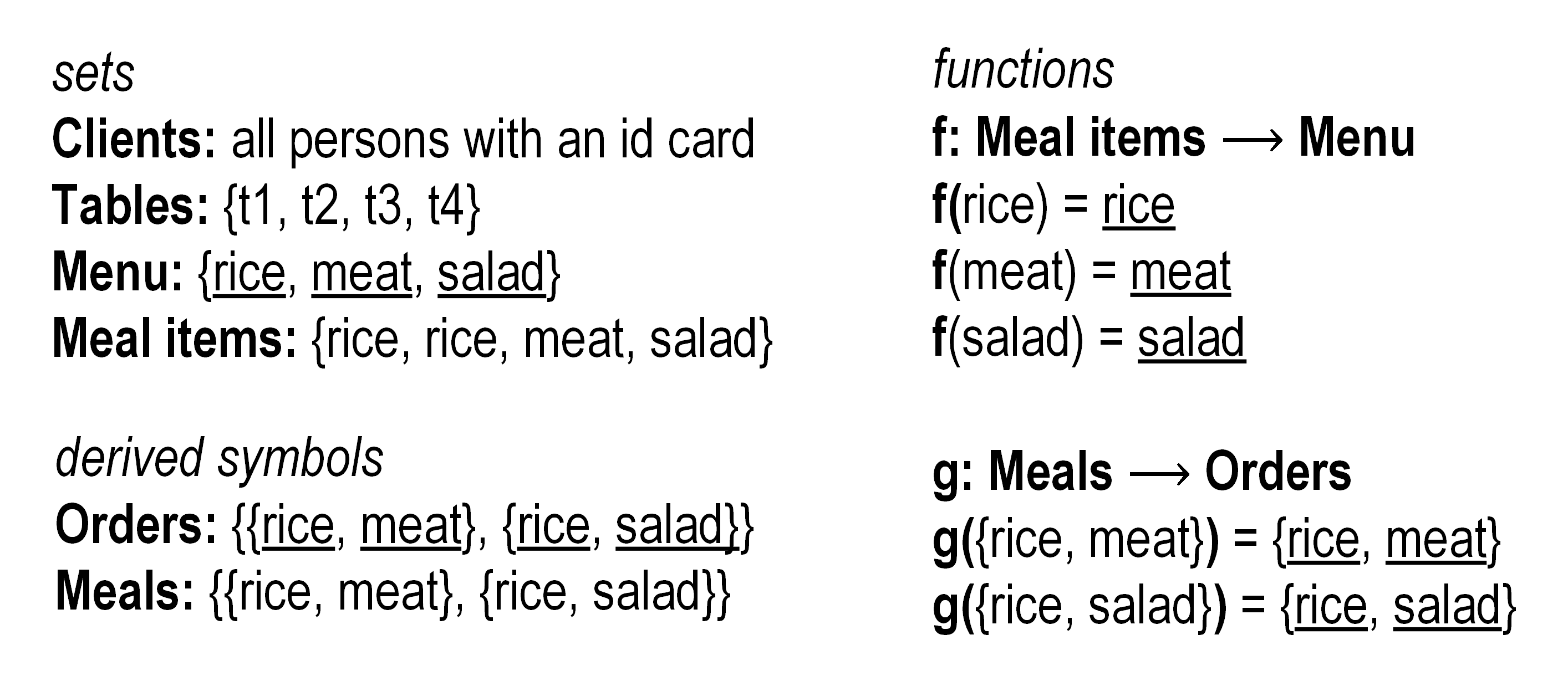}
\caption{$\Sigma$-Struktur $S_0$}
\label{fig:Struktur}
\end{minipage}
\end{figure}

Um dennoch für alle Filialen die Abläufe gleich zu formulieren, verwenden wir eine Signatur, $\Sigma_0$, die Abb.~\ref{fig:Signatur} zeigt. Sie enthält die vier Symbole \textit{Clients, Tables, Menu, Meal items} für die vier oben beschriebenen Mengen, dazu zwei Symbole (\textit{Orders} und \textit{Meals}) für Teilmengen und zwei Symbole ($f$ und $g$) für Funktionen. \textit{Orders} steht für Bestellungen (eine Bestellung ist eine Teilmenge der Einträge in die Speisekarte), und \textit{Meals} steht für Gerichte (ein Gericht ist eine Teilmenge zubereiteter Speisen). Die Funktion $f$ ordnet jeder Speise den entsprechenden Eintrag in der Speisekarte zu; $g$ ordnet jedem Gericht seine Bestellung zu.

In Abschnitt 3.2 wurde diskutiert, wie eine Signatur auf vielfältige Weise instanziiert werden kann. Jede konkrete Filiale ist durch eine solche Instanziierung der Signatur $\Sigma_0$ charakterisiert. In der Fallstudie diskutieren wir die Instanziierung $S_0$ der Abb.~\ref{fig:Struktur}.

\subsection{Das Verhalten der drei Module auf Schema-Ebene}

\begin{figure}[!tb]
\centering
\includegraphics[scale=.20]{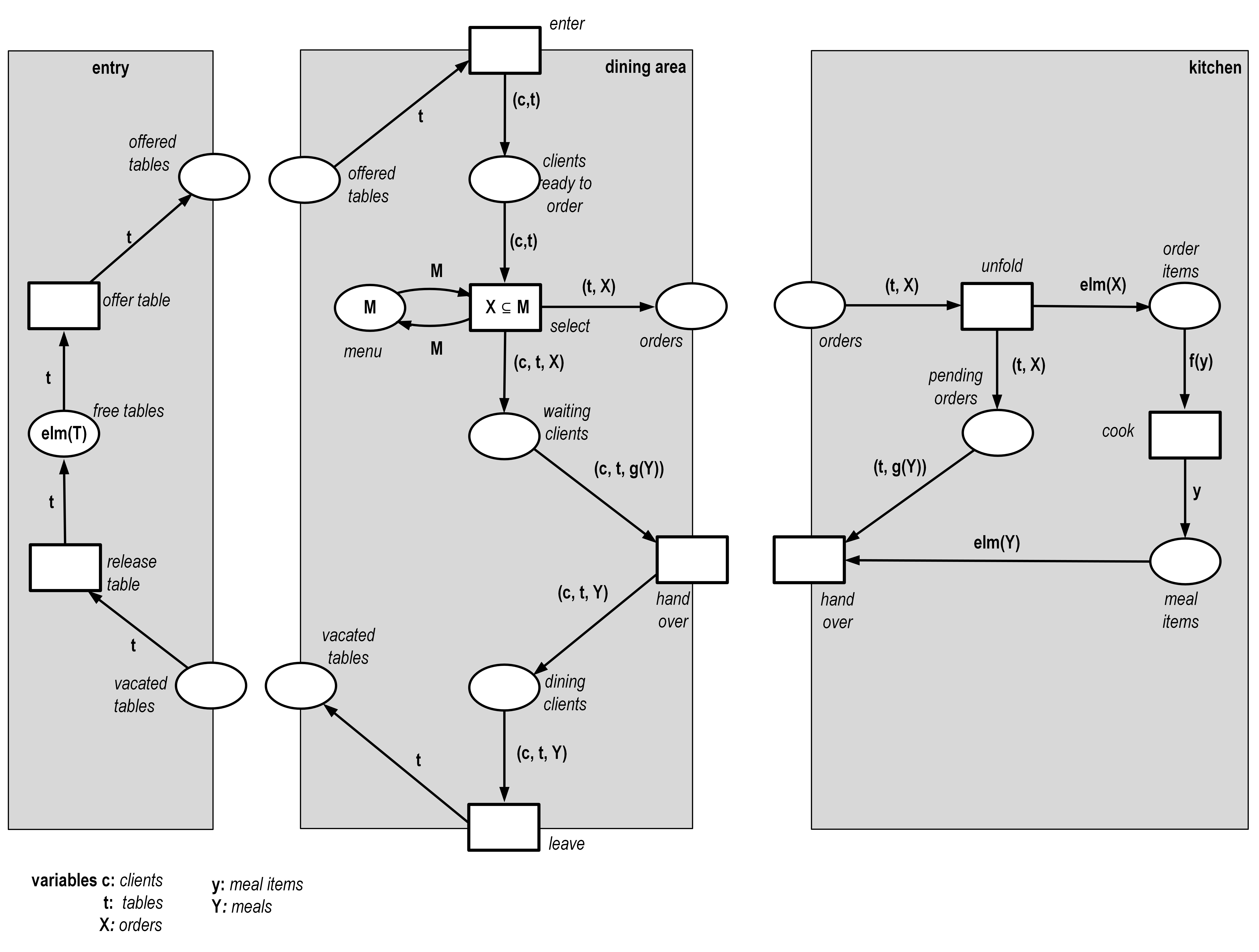}
\caption{Verhalten der Module auf der Schema-Ebene}
\label{fig:Schema-Module}
\end{figure}

Zunächst modellieren wir das Verhalten jedes einzelnen der drei Module aus Abb.~\ref{fig:drei_Module}. Abb.~\ref{fig:Schema-Module} zeigt das Verhalten der einzelnen Module auf der Schema-Ebene, also auf Basis der Signatur $\Sigma_0$, und nicht einer einzelnen Instanziierung.  Als mathematisches Konzept repräsentiert ein Platz ein Prädikat, das auf eine Menge von Objekten zutrifft. Diese Menge kann durch den Eintritt von Transitionen wachsen und schrumpfen.

Das \textit{entry}-Modul in Abb.~\ref{fig:Schema-Module} verwaltet die Tische.  Zunächst stellt sich hier die Frage nach einer angemessenen Anfangsmarkierung: Für eine gegebene Instanziierung, also eine konkrete Filiale, enthält der Platz \textit{free tables} anfangs alle Tische als Marken. Im Beispiel der Struktur $S_0$ aus Abb.~\ref{fig:Struktur} sind das die vier Marken $t_1, t_2, t_3, t_4$. Auf der schematischen Ebene kennen wir nur das Symbol \textit{Tables}, für das jede Instanziierung eine Belegung mit einer Menge von Tischen frei wählen kann.  Die zunächst naheliegende Idee, den Platz \textit{free tables} anfangs mit dem Symbol \textit{Tables} zu beschriften, greift allerdings zu kurz: Bei einer Instanziierung des Symbols \textit{Tables} mit einer Menge, beispielsweise $\{t_1, t_2, t_3, t_4\}$, würde anfangs diese Menge als eine einzige Marke entstehen. Wir wollen aber die vier Elemente dieser Menge als einzelne Marken. Das wird mit \textit{elm(Tables)} notiert.

Mathematisch steckt dahinter, dass ein Platz ein Prädikat darstellt, das aktuell auf die Elemente im Platz zutrifft. Der Platz \textit{free tables} mit der Inschrift \textit{elm(Tables)} steht also für den logischen Ausdruck:
\begin{equation}
    \forall t \in \textit{Tables}: \textit{free tables(t)}
\end{equation}
Die Instanziierung $S_0$ erzeugt damit in der Anfangsmarkierung den logischen Ausdruck „$\forall t \in \{t_1, t_2, t_3, t_4\}: \textit{free tables(t)}$“. Damit liegen auf dem Platz \textit{free tables} anfangs die vier Marken $t_1, t_2, t_3, t_4$.

In einer gegebenen Instanziierung $S$ kann im Gastraum-Modul die Transition \textit{enter} eintreten, sobald eine Marke auf \textit{offered tables} vorliegt, indem die Variable $t$ mit dieser Marke belegt wird. Die Belegung der Variablen $c$ kann frei gewählt werden aus der Menge, mit der die Instanziierung $S$ das Mengensymbol \textit{Clients} instanziiert. Die Idee dahinter: Das Modul kann mit einem anderen Modul aus seiner Umgebung komponiert werden, indem die Belegung von $c$ mit einem Gast einen inhaltlichen Sinn hat. 

Bei der Transition \textit{select} ist das Verhältnis zwischen dem Symbol \textit{Menu} und der Variablen $X$ interessant:  Eine Instanziierung $S$ des Gastraum-Moduls legt \textit{Menu} als eine Speisekarte, also eine Menge einzelner Einträge fest. Hingegen wird beim Eintreten von \textit{select} die Variable $X$ jedes mal neu belegt. Die Bedingung $X \subseteq \textit{Menu}$ der Inschrift von \textit{select} garantiert, dass die Belegung von $X$ („Bestellung“) nur Einträge enthält, die in der Speisekarte aufgeführt sind. Eine \textit{order} besteht aus einem Tisch und einer Bestellung. Jede Marke auf dem Platz \textit{waiting} besteht aus einem Kunden $c$, seinem Tisch $t$, und seiner Bestellung $X$. Mit der Transition \textit{hand over} wird dem Gast das bestellte Gericht ausgehändigt.

Im \textit{kitchen}-Modul zerlegt die Transition \textit{unfold} jede eingehende Bestellung in ihre einzelnen Einträge, formuliert mit Hilfe des \textit{elm}-Operators, analog zur Verwendung im \textit{entry}-Modul. Für jeden dieser Einträge $f(y)$, also jede Benennung einer Speise, wird die Speise $y$ gekocht (Transition \textit{cook}). Entsprechend einer vorliegenden Bestellung auf dem Platz \textit{pending orders} werden dann Speisen als $Y$ zusammengestellt und als Gericht an den Kunden übergeben.

\subsection{Die Instanziierung $S_0$ und ihre Verhalten}
\begin{figure}[!tb]
\centering
\includegraphics[scale=.25]{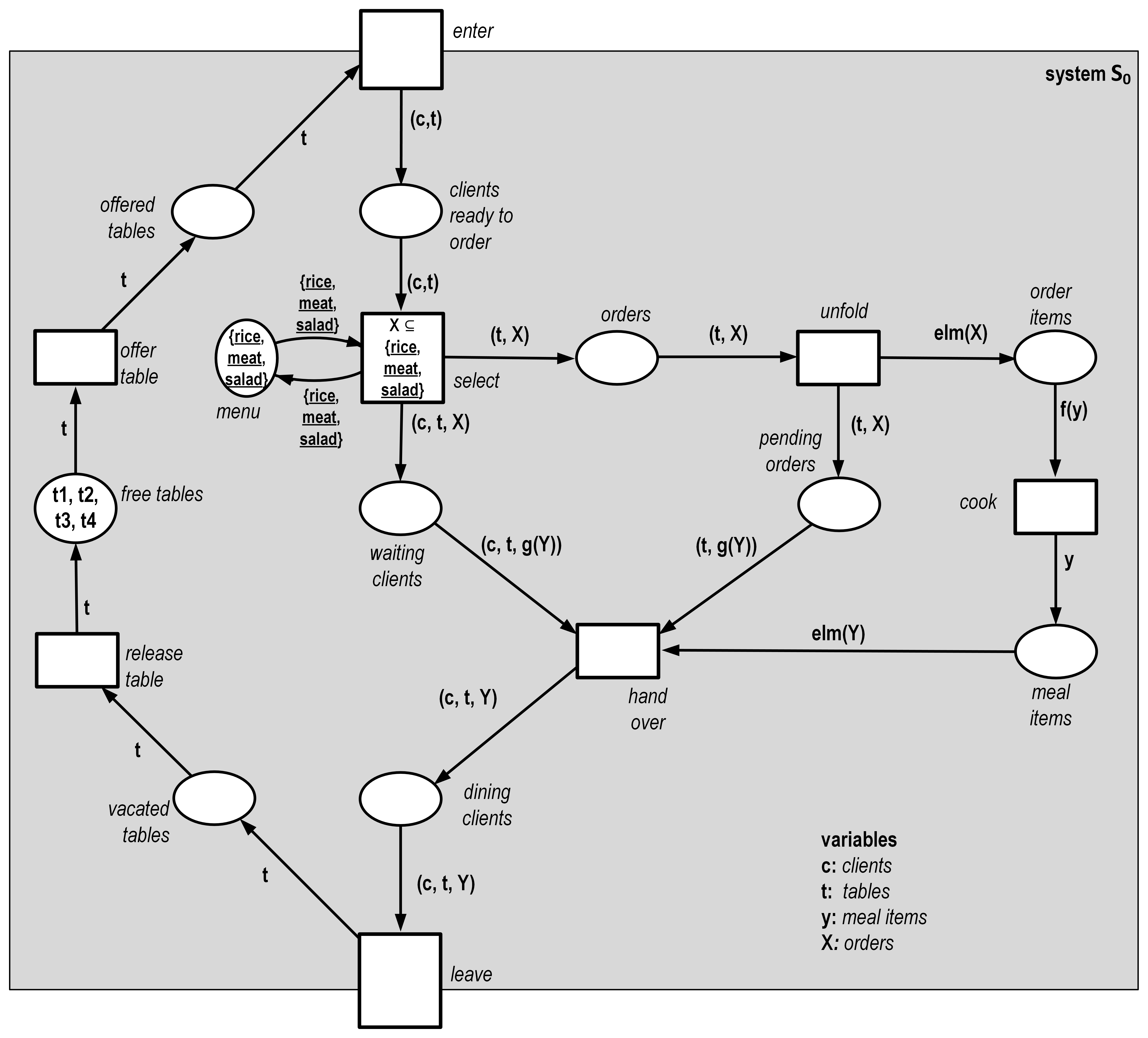}
\caption{System $S_0$}
\label{fig:System_S0}
\end{figure}

Abb.~\ref{fig:System_S0} zeigt ein Modul, $S_0$, das aus den Modulen von Abb.~\ref{fig:Schema-Module} in zwei Schritten entsteht: erstens werden die drei Module aus Abb.~\ref{fig:Schema-Module} zu einem einzigen Modul, System $S_0$,  komponiert; zweitens wird die Signatur $\Sigma_0$ von Abb.~\ref{fig:Schema-Module} mit der Struktur $S_0$ instanziiert. Insbesondere enthält der Platz \textit{free tables} jetzt vier Marken, und der Platz \textit{menu} drei Marken.

In der Instanziierung $S_0$ kann man nun den Eintritt einzelner Transitionen dokumentieren. Im Anfangszustand kann beispielsweise die Transition \textit{offer table} mit der Belegung $t = t_1$ eintreten und dabei die Marke $t_1$ auf den Platz \textit{offered tables} verschieben. Danach kann \textit{enter} mit der Belegung $t = t_1$ und einem frei gewählten Kunden für $c$ eintreten. Unabhängig davon kann \textit{offer table} auch mit $t = t_2$ eintreten, etc. So kann beispielsweise ein Ablauf entstehen, in dem der Gast \textit{Alice} aus der Speisekarte eine Bestellung für Reis und Fleisch zusammenstellt, für die in der Küche die entsprechenden Speisen gekocht und an Alice übergeben werden. Daneben kann \textit{Bob} ein Gericht mit Reis und Salat bestellen und ausgehändigt bekommen. Solches einzelnes Verhalten wird üblicherweise als Sequenz aus globalen Zuständen und eintretenden Transitionen beschrieben, beginnend mit der Anfangsmarkierung. 
 
\textsc{Heraklit} modelliert ein einzelnes Verhalten als verteilten Ablauf (vgl. Abschnitt 3.3); formal gefasst als ein Modul. Damit können verteilte Abläufe einfach komponiert werden. Am laufenden Beispiel wird zudem deutlich, dass verteilte Abläufe Einzelheiten der Zusammenhänge innerhalb eines Ablaufs sehr viel expliziter zeigen als es mit sequentiellen Abläufen möglich ist. 

\begin{figure}[!tb]
\centering
\includegraphics[scale=.20]{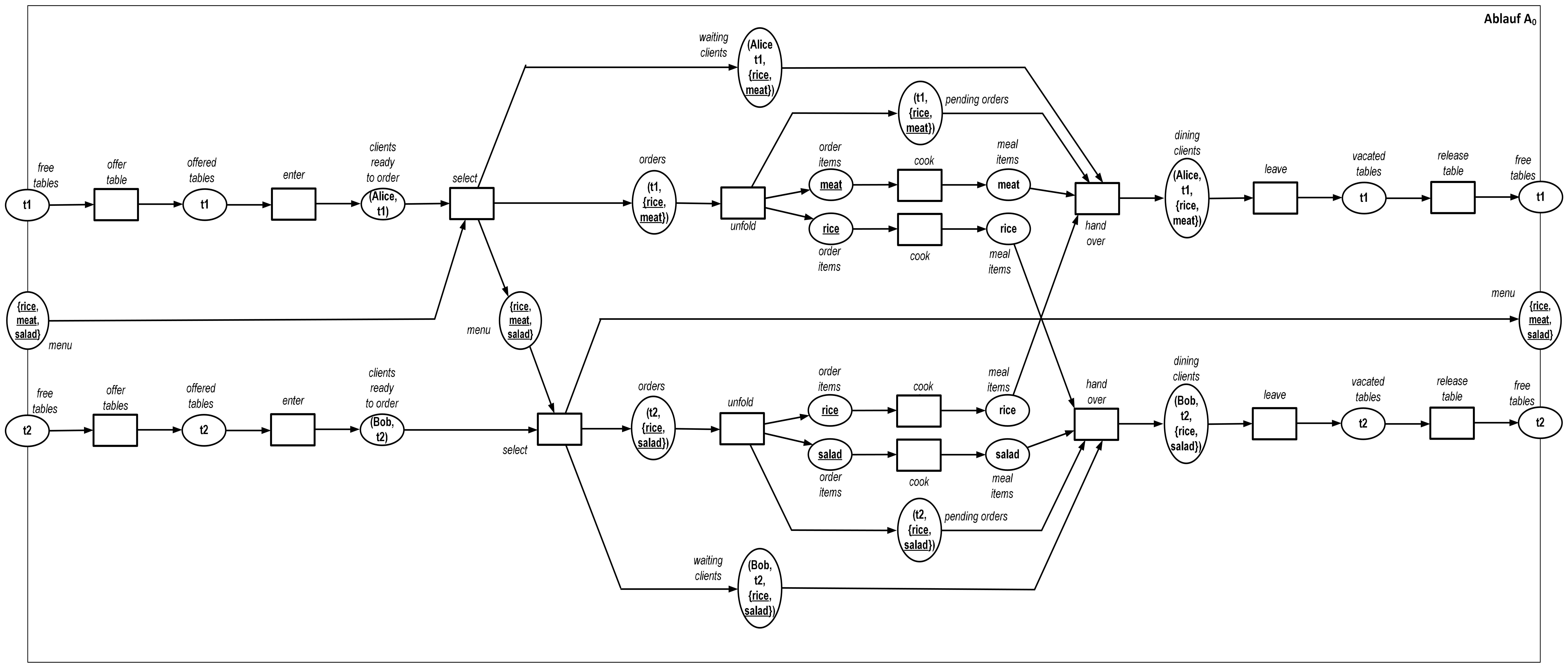}
\caption{Ablauf $A_0$}
\label{fig:Ablauf_gesamt}
\end{figure}

\begin{figure}[!tb]
\begin{minipage}{.5\textwidth}
\centering
\includegraphics[scale=.21]{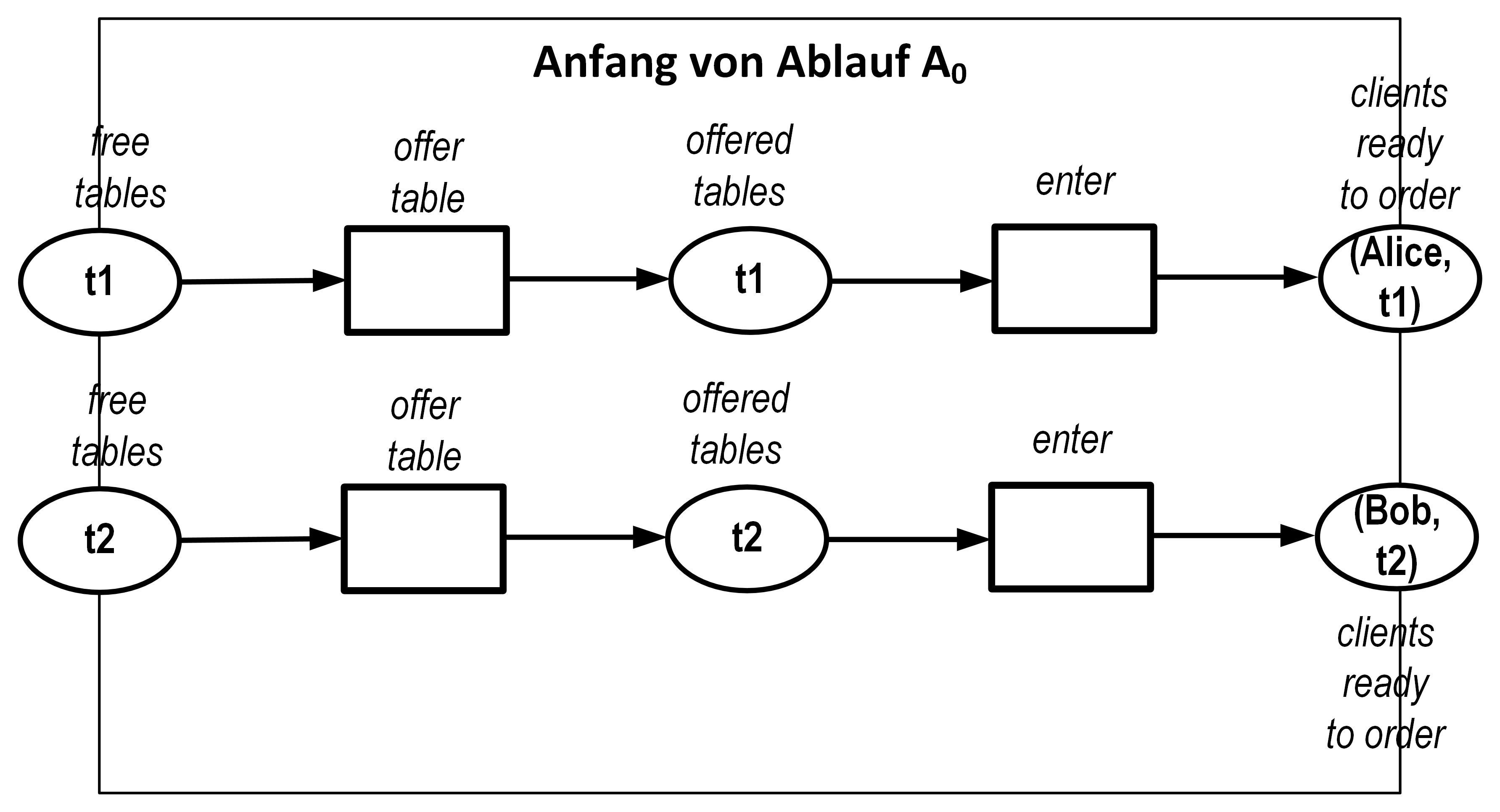}
\caption{Anfangsstück des Ablaufs $A_0$}
\label{fig:Ablauf_Teil_1}
\end{minipage}%
\begin{minipage}{.5\textwidth}
\centering
\includegraphics[scale=.21]{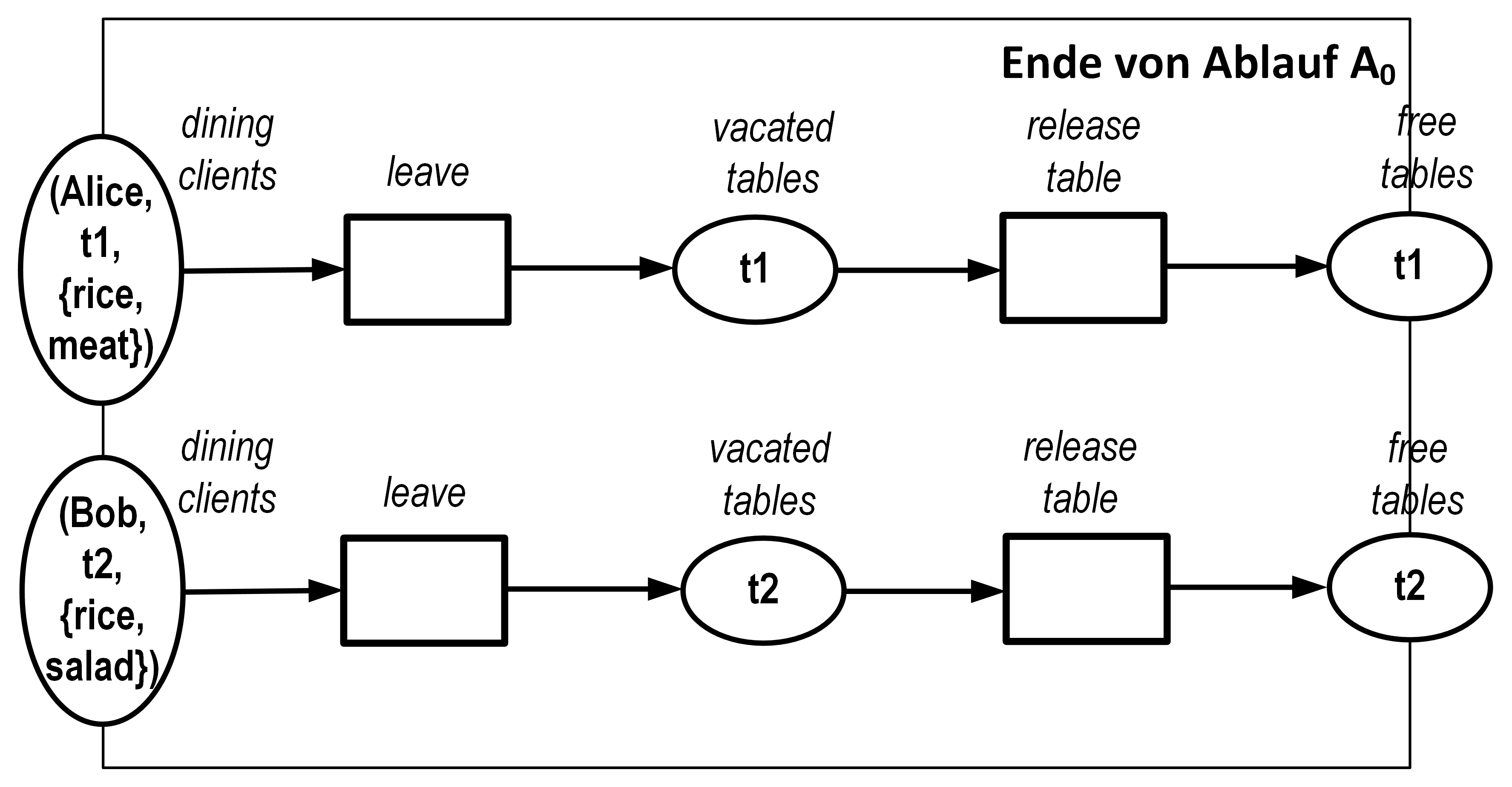}
\caption{Endstück des Ablaufs $A_0$}
\label{fig:Ablauf_Teil_3}
\end{minipage}
\end{figure}

Abb.~\ref{fig:Ablauf_gesamt} zeigt einen verteilten Ablauf, $A_0$, des Systems $S_0$ in Abb.~\ref{fig:System_S0}. Zur besseren Lesbarkeit komponieren wir diesen Ablauf aus drei Teilen: Das Modul \textit{Anfang von $A_0$} in Abb.~\ref{fig:Ablauf_Teil_1} besteht aus zwei Strängen. Der obere beginnt mit dem Eintritt der Transition \textit{offer table} aus Abb.~\ref{fig:System_S0} im Modus $t = t_1$. Dadurch enthält der Platz \textit{offered tables} die Marke $t_1$. Diese Marke wiederum aktiviert die Transition \textit{enter} mit der Belegung $t = t_1$, und einer frei wählbaren Belegung von $c$. Der Ablauf $A_0$ wählt $c = \textit{Alice}$ und erzeugt damit die Marke $(\textit{Alice}, t_1)$ auf dem Platz \textit{clients ready to order}. Unabhängig von diesem Strang beschreibt der untere Strang Schritte der Marke $t_2$ und das Entstehen der Marke \textit{$(\textit{Bob}, t_2)$} auf dem Platz \textit{clients ready to order}.

\begin{figure}[!tb]
\centering
\includegraphics[scale=.21]{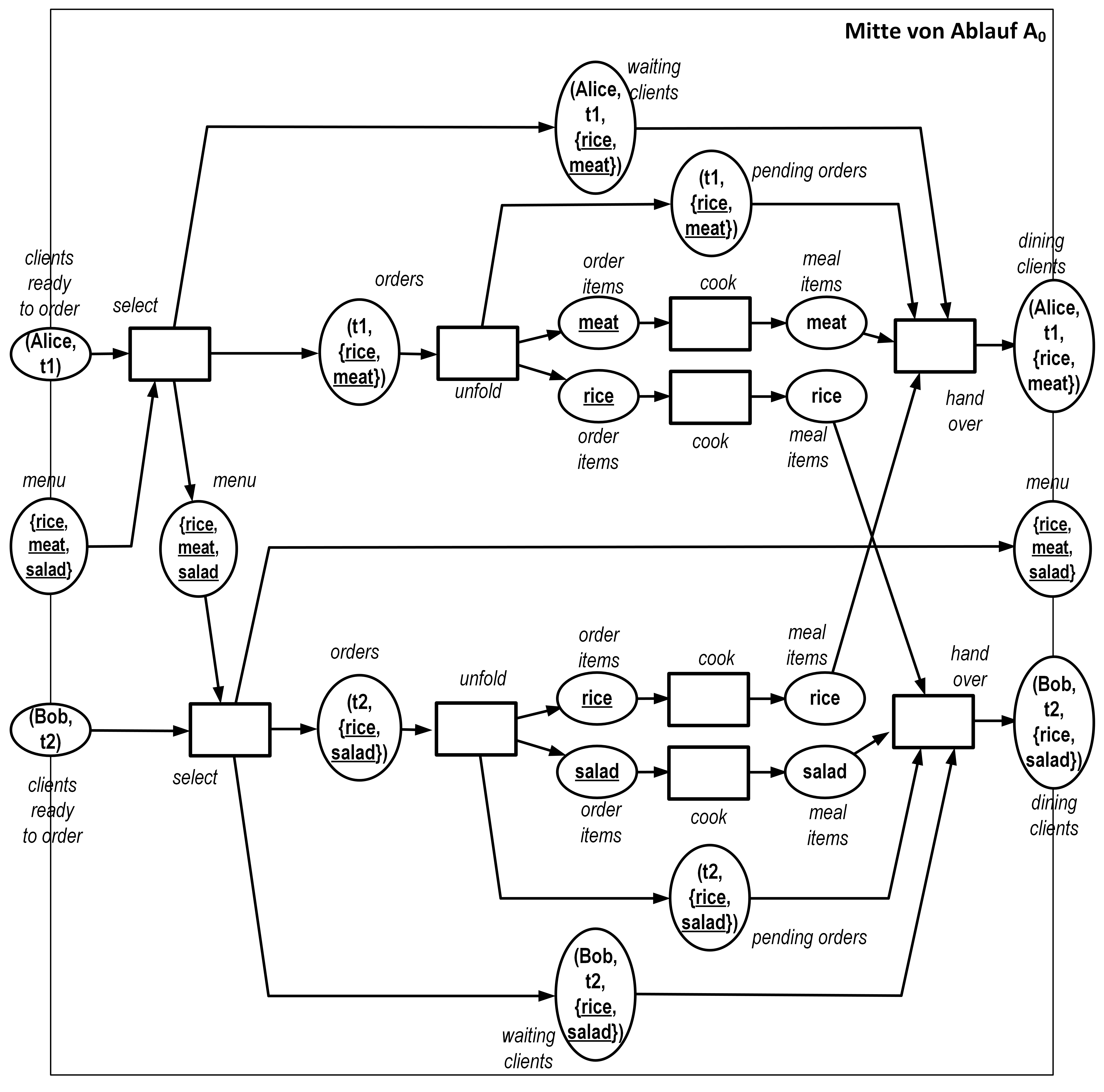}
\caption{Mittelstück des Ablaufs $A_0$}
\label{fig:Ablauf_Teil_2}
\end{figure}

Das Modul \textit{Mitte von $A_0$} in Abb.~\ref{fig:Ablauf_Teil_2} ergänzt in seiner linken Schnittstelle die rechte Schnittstelle von \textit{Anfang von $A_0$} um den Platz \textit{menu} mit der Marke \textit{\{rice, meat, salad\}}. Diese Marke repräsentiert das einzige verfügbare Exemplar der Speisekarte. \textit{Alice} und \textit{Bob} schauen sie nacheinander an, und erzeugen jeweils eine \textit{order}, $(t_1, \{\underline{rice}, \underline{meat}\})$ und $(t_2, \{\textit{\underline{rice}}, \textit{\underline{salad}}\})$. An der Transition \textit{unfold} beginnt im Modul $\textit{System } S_0$ ein Pfeil mit der Inschrift $\textit{elm}(X)$. Mit $t = t_1$ und $X = \{\textit{\underline{rice}}, \textit{\underline{meat}}\}$  zerlegt $\textit{elm}(X)$ im Modul den Anteil $\{\textit{\underline{rice}}, \textit{\underline{meat}}\}$  der Marke $(t_1, \{\underline{rice}, \underline{meat}\})$ in die Bestandteile \textit{\underline{rice}} und \textit{\underline{meat}}, analog zur Inschrift des Platzes \textit{free tables} im Schema von Abb.~\ref{fig:Schema-Module}. Die entsprechenden Speisen \textit{rice} und \textit{meat} werden einzeln gekocht. Mit der Bestellung des Tisches $t_2$ wird entsprechend verfahren. In beiden Fällen wird \textit{rice} bestellt. Da die Bestellungen nebenläufig vorliegen, können die beiden gekochten \textit{rice}-Portionen nicht eindeutig den Bestellungen zugeordnet werden. Im Ablauf $A_0$ werden sie vertauscht.

Das Modul \textit{Ende von Ablauf $A_0$} in Abb.~\ref{fig:Ablauf_Teil_3} beendet nun die beiden Ablauf-Teile in offensichtlicher Weise. Die Komposition \textit{Anfang von Ablauf} $A_0 \bullet \textit{Mitte von } A_0 \bullet \textit{Ende von Ablauf } A_0$ der drei Module liefert genau das Modul $\textit{Ablauf Ablauf } A_0$ in Abb.~\ref{fig:Ablauf_gesamt}.

\section {Bezug von \textsc{Heraklit} zu anderen Modellierungsinfrastrukturen}
Ein \textit{Framework} zur Modellierung beschreibt wesentliche Konzepte zur Modellierung. Darauf aufbauend bietet eine \textit{Modellierungsinfrastruktur} korrespondierende Modelle, Methoden, Techniken und Werkzeugen. Im Vergleich mit anderen Disziplinen wird generell \textit{in der Praxis} der (Wirtschafts-) Informatik wenig mit Modellen gearbeitet. Es sind zahlreiche Frameworks vorgeschlagen worden, aber keines hat sich bisher wirklich durchgesetzt.

Ein Beispiel sind Diagramme der \textit{Unified Modeling Language} (UML, \cite{OMG2017UML}) für Software-Systeme. Die verschiedenen Diagrammtypen veranschaulichen einige Aspekte eines Softwaresystems, aber unterstützen Analysefragen nur wenig. Ein anderes Beispiel sind die in der Wirtschaftsinformatik beliebten Diagramme der \textit{Business Process Model and Notation}
(BPMN, \cite{OMG2014BPMN}) oder \textit{ereignisgesteuerte Prozeßketten} (EPK, \cite{keller1992EPK}). Solche Diagramme beschränken sich auf die Identifikation abstrakter Aktivitäten und die Darstellung des Kontrollflusses. Sie unterstützen Abstraktion und Komposition, helfen aber wenig beim Umgang mit konkreten oder abstrakten Daten und Gegenständen bei der Beschreibung des Verhaltens und beim Nachweis der Korrektheit von Verhaltensmodellen gegenüber einer Spezifikation. Die \textit{Architektur integrierter Informationssysteme} (ARIS, \cite{scheer2001ARIS}) erlaubt zwar eine integrierte Beschreibung von Daten, Funktionen und Abläufen. Allerdings fehlen leistungsfähige Modularisierungskonzepte. Auch sind ARIS-Modelle nur bedingt formalisiert. Weit verbreitet sind Petrinetze; in ihrer klassischen Form modellieren sie allerdings lediglich verteilten Kontrollfluss. Zahlreiche Varianten und Verallgemeinerungen (insbesondere \textit{Coloured Petri Nets} (CPN, \cite{jensen2009colourednets})) integrieren konkrete Daten (in der Sprechweise von \textsc{Heraklit}: einzelne Strukturen). Typische Vorschläge für hierarchische Petrinetze ersetzen einzelne Transition durch ganze Netze. Statecharts \cite{harel1987statecharts} verwenden eigenständige Modul- und Kompositionskonzepte; dabei werden gleich gelablete Übergänge endlicher Automaten verschmolzen. \cite{graics2020mixed} schlagen die statecart-basierte Sprache \textit{Gamma} vor und diskutieren eine Vielzahl von Kompositionsoperatoren.

Schließlich gibt es noch eine Reihe von Frameworks, die letztlich als Strukturen einer Signatur, oder als eine Signatur auffassbar sind. Dazu gehören Abstract State Machines (ASM, \cite{gurevich2000ASM}), event-B \cite{abrial2005B} und Z \cite{spivey1992B}. Andere Frameworks orientieren sich an der Prädikatenlogik, beispielsweise TLA \cite{lamport2002TLA}, FOCUS \cite{broy1997refinement,broy2001specification} und Aloy \cite{jackson2019alloy}.

Alle diese Frameworks werden zur Analyse und Simulation von Softwarewerkzeugen unterstützt und wurden in größeren Software-Entwicklungsprojekten eingesetzt. Allerdings hat sich keine wirklich durchgesetzt. Das liegt vorwiegend daran, dass bisher aus Modellen nicht besonders viel Nutzen gezogen werden kann. Derzeitige Frameworks fokussieren jeweils nur spezielle Bereiche eines Systems. Für kleine und mittelgroße Systeme sind die genannten Frameworks mehr oder weniger hilfreich. Aber da, wo man es besonders braucht, bei wirklich großen Systemen, fehlen systematische, strukturierende und abstrahierende Prinzipien zur Modellierung. Komplexe Systeme werden bestenfalls in Teilen modelliert, oft mit ganz unterschiedlichen Frameworks, die nur mühsam zusammenpassen. Es gibt keine umfassenden Modellsichten, in die dann ganz unterschiedliche Teilmodelle integriert werden können.

Im Vergleich mit den genannten Frameworks ist  \textsc{Heraklit} breiter aufgestellt, indem  Petrinetze, Konzepte algebraischer Spezifikation, und ein universeller Kompositionsoperator miteinander integriert sind. Keine der erwähnten Frameworks erreicht die Ausdruckskraft von \textsc{Heraklit}. Mit keiner der erwähnten Frameworks können alle Aspekte der Fallstudie des vierten Abschnitts modelliert werden.

Mit dem in der Logik grundlegenden und in der algebraischen Spezifikation bewährten Konzept einer Signatur und ihren unterschiedlichen Instanziierungen beschreibt \textsc{Heraklit} auch dynamisches Verhalten, sowohl auf der schematischen Ebene als auch für einzelne Instanziierungen. Das war auch die Intention von Prädikat-Transitionsnetzen \cite{genrich1981predicatenets}.

\textsc{Heraklit} legt Wert auf Ausdrucksmittel für eine integrierte Modellierung lebensweltlicher und formaler Sachverhalte. Dazu nutzt \textsc{Heraklit} die wissenschaftstheoretischen Diskussionen um den Bezug zwischen der Welt und ihrer formalen Fassung, seit Aristoteles, die dann in die Prädikatenlogik eingeflossen ist \cite{suppes1957logic}. \textsc{Heraklit} schlägt vor, dynamische Aspekte in die Prädikatenlogik einzubauen.

Wie bereits in der Einleitung erwähnt, hat \textit{Dijkstra} vorgeschlagen, informelle und formale Argumentationen strikt zu trennen und zwischen ihnen eine „Firewall“ zu errichten.  Diese Mauer ist nicht hilfreich zum Verständnis Informatik-basierter Systeme. Vielmehr braucht es eine Brücke, um zwischen der informellen Lebenswelt der Anwendung und der formalen Welt der Technik zu vermitteln. \textsc{Heraklit} bietet hierzu ein passendes Framework. 

\section{Ausblick} Die formalen Grundlagen und die Prinzipien des Einsatzes von \textsc{Heraklit} liegen vor; zahlreiche Fallstudien zeigen den Nutzen des umfassenden Ansatzes von \textsc{Heraklit}. Für die Verwendung im industriellen Maßstab und als bessere Alternative zu derzeit verwendeten Frameworks zur Modellierung fehlen insbesondere noch Softwarewerkzeuge zur Unterstützung des Entwurfs großer Modelle, und auf \textsc{Heraklit} zugeschnittene Analyseverfahren. Für spezielle Anwendungsbereiche werden im Lauf der Zeit Ausprägungen mit verfeinerten Konzepten gebildet. In einigen Anwendungen ist auch die automatische Erzeugung von Programmcode für einige Module wünschenswert.

\bibliography{main} 
\end{document}